\documentclass[english,nofootinbib,superscriptaddress]{revtex4-1}
\usepackage{lmodern}

\usepackage[T1]{fontenc}
\usepackage[latin9]{inputenc}
\usepackage[letterpaper]{geometry}
\usepackage{amsmath}
\usepackage{amssymb}
\usepackage{bbm}
\usepackage{babel}
\usepackage{graphicx}

\begin{document}
\title{Speed of particles and a relativity of locality\\
in $\kappa$-Minkowski quantum spacetime}

\author{Giovanni AMELINO-CAMELIA}
\affiliation{\footnotesize{Dipartimento di Fisica, Universit\`a di Roma ``La Sapienza", P.le A. Moro 2, 00185 Roma, EU}}
\affiliation{\footnotesize{INFN, Sez.~Roma1, P.le A. Moro 2, 00185 Roma, EU}}
\author{Niccol\'o LORET}
\affiliation{\footnotesize{Dipartimento di Fisica, Universit\`a di Roma ``La Sapienza", P.le A. Moro 2, 00185 Roma, EU}}
\affiliation{\footnotesize{INFN, Sez.~Roma1, P.le A. Moro 2, 00185 Roma, EU}}
\author{Giacomo~ROSATI}
\affiliation{\footnotesize{Dipartimento di Fisica, Universit\`a di Roma ``La Sapienza", P.le A. Moro 2, 00185 Roma, EU}}
\affiliation{\footnotesize{INFN, Sez.~Roma1, P.le A. Moro 2, 00185 Roma, EU}}

\begin{abstract}
\noindent
The last decade of research on $\kappa$-Minkowski noncommutative spacetime
has been strongly characterized by a controversy concerning
the speed of propagation of massless particles. Most arguments suggested that
this
speed should depend on the momentum of the particle
strongly enough to be of interest for some ongoing experimental studies.
But the only explicit derivations
of worldlines in $\kappa$-Minkowski predicted no momentum dependence for the speed
of massless particles. We return to this controversy
equipped with the recent understanding
(arXiv:1006.2126, arXiv:1007.0718, arXiv:1008.2962,
arXiv:1101.0931) that in some quantum spacetimes
coincidences of events assessed by an observer who is distant from
the events can be artifactual, and particularly Smolin's thesis (arXiv:1007.0718)
that $\kappa$-Minkowski should be an example of such a spacetime.
We therefore set up our investigation in such a way that we never rely on
the assessment of coincidences of events by distant observers.
This allows us to verify explicitly that in $\kappa$-Minkowski
simultaneously-emitted massless particles of different momentum
are detected at different times, and establish a linear dependence of the
detection times on momentum.
\end{abstract}

\maketitle

\section{Introduction}
Over the last decade there has been a strong effort aimed at seeking
experimental evidence of Planck-scale features of the quantum-gravity
and/or quantum-spacetime realm~\cite{gacLRR}.
One of the most studied opportunities concerns the possibility
that the speed of massless particles (photons) might have a Planck-scale-induced
dependence on wavelength/momentum. On the theory side this finds motivation in
several semi-heuristic but compelling analyses of quantum-gravity/quantum-spacetime
frameworks which indeed expose plausible mechanisms for such a momentum dependence
to arise~\cite{aemn1,grbgac,gampul,mexweave,gacmaj}.
And from the viewpoint of experimental tests it is noteworthy that
some of these theory scenarios produce effects that are within the range
of sensitivities of analyses exploiting data presently gathered
with powerful gamma-ray
telescopes~\cite{grbgac,fermiSCIENCE,ellisUNO,gacsmolinPRD,fermiNATURE}.

While the opportunity to look for effects motivated by work on some
quantum-gravity/quantum-spacetime theories is certainly exciting,
for the assessment of the possible impact of these studies on the development
of quantum gravity it is crucial that we establish whether any models
could be falsified.
If we stumble upon positive/discovery experimental results it would evidently
be a huge impulse for quantum-gravity research. But if it turns out
 that the results are negative (which of course is most likely)
 will we have learned anything
substantial about the quantum-gravity problem? Would ideed some models be falsified?\\
From this perspective it remains of paramount importance to establish
that the momentum dependence of the speed of massless particles is
a definite prediction (rather than a plausible outcome) of some specific
models of quantum gravity and/or quantum spacetime.
The most promising opportunity of this type is found in studies
of the $\kappa$-Minkowski quantum
spacetime~\cite{majrue,zakKAPPAMINK,lukieANNALS,gacmaj},
where several lines of analysis appear to be very close to a fully rigorous
derivation of a result on the momentum dependence of the speed of massless particles.
As a matter of fact we might already have such a robust result, the one reported
in Ref.~\cite{gacmaj}, exploiting the properties of the $\kappa$-Minkowski
quantum differential calculus to derive constructively the speed of
propagation of classical $\kappa$-Klein-Gordon waves.
The main obstruction for adopting this momentum-dependence as
a ``consensus result" of the $\kappa$-Minkowski literature
is found within an approach based on ``$\kappa$-Minkowski phase spaces",
 where one can derive wordlines of
massless particles, apparently finding no such momentum dependence~\cite{jurekvelISOne}.

Because of its relevance both for phenomenology and from a theory perspective
this issue of the description of the speed of massless particles in $\kappa$-Minkowski
has been hotly debated
(see, {\it e.g.},
Refs.~\cite{gacmaj,lukieANNALS,jurekvelISOne,kowaVEL1,japaVEL,lukieVEL,gacMandaniciDANDREA,kosinskiVEL,mignemiVEL,tezukaVEL,grilloVEL,majid2006,ghoshVELisONE,ghoshVEL}).
 We return to this issue
equipped with the recent
understanding~\cite{whataboutbob,leeINERTIALlimit,arzkowaRelLoc,principle,laurentPRL}
that in some quantum spacetimes
coincidences of events assessed by an observer who is distant from
the events can be artifactual, and particularly Smolin's thesis~\cite{leeINERTIALlimit} that $\kappa$-Minkowski should be an example of such a spacetime.
We handle this ``relativity of locality" by
 setting up our investigation
 of worldlines in $\kappa$-Minkowski
 in such a way that we never rely on
the assessment of coincidences of events by distant observers.
We find that in $\kappa$-Minkowski 
simultaneously-emitted massless particles of different momentum
are detected at different times, and establish a linear dependence of the
detection times on momentum, in full agreement with the previous
independent derivation reported in Ref.~\cite{gacmaj}.

\section{Time-to-the-right formulation of $\kappa$-Minkowski}
We shall focus throughout on the most studied formulation
of theories in $\kappa$-Minkowski spacetime, often labeled
as ``bicrossproduct basis"~\cite{majrue,lukieANNALS}
 or ``time-to-the-right basis"~\cite{gacAlessandraFrancesco}.

We adopt the
convention $\eta_{\mu\nu} =\{ -1,1,1,1\}$
for the Lorentzian metric,
and we adopt units such that the speed-of-light scale
(speed of massless particles in the infrared limit)
and the reduced Planck constant
are $1$ ($c=1=\hbar$).

 Starting from $\kappa$-Minkowski noncommutativity~\cite{zakKAPPAMINK,majrue,lukieANNALS}
  \begin{equation}
\left[ x_j,x_0 \right] = i\ell x_j ~ ,
\label{kappadefCOMMUT}
\end{equation}
in this formulation one introduces the Fourier transform $\tilde{\Phi}(k)$
of a given $\kappa$-Minkowski field $\Phi(x)$ using the
time-to-the-right convention
$$\Phi(x) = \int d^4x\tilde{\Phi}(k)e^{ik_jx^j}e^{ik_0x^0}~. $$
This is often equivalently described in terms of the time-to-the-right
Weyl map ${\cal W}_R$
by writing that
$$\Phi(x) = {\cal W}_R \left(\int d^4x\tilde{\Phi}(k)e^{ik_\mu x^\mu} \right)$$
where it is intended that coordinates trivially commute when placed
inside the Weyl map, ${\cal W}_R (x_j x_0) = {\cal W}_R (x_0 x_j)$,
and that taking a function out of the time-to-the-right Weyl map
implies~\cite{gacAlessandraFrancesco}
time-to-the-right ordering, so that
for example $x_j x_0 = {\cal W}_R (x_j x_0)$, $x_j x_0 = {\cal W}_R (x_0 x_j)$,
and $e^{ik_jx^j}e^{ik_0x^0} = {\cal W}_R (e^{ik_\mu x^\mu})$.

Then several arguments~\cite{majrue,lukieANNALS}, including the ones
based on the  recently-developed techniques of Noether analysis~\cite{kappanoether,freidkowaNOETHER,nopure},
lead one to find generators of symmetries under
translations, space-rotations and boosts.
For translations one has that
  \begin{equation}
P_\mu e^{ik_jx^j}e^{ik_0x^0} = k_\mu e^{ik_jx^j}e^{ik_0x^0}=
{\cal W}_R \left( - i \partial_\mu e^{ik_\nu x^\nu} \right) ~ ,
\label{traslCOMMUT}
\end{equation}
and in general $P_\mu {\cal W}_R \left( f(x) \right) =
{\cal W}_R \left( - i \partial_\mu f(x) \right)$.\\
Similarly one has that the generators of space rotations
are given by
$$R_l e^{ik_jx^j}e^{ik_0x^0} = \epsilon_{lmn} x_m k_n e^{ik_jx^j}e^{ik_0x^0} $$
and the generators of boosts are given by
$$\mathcal{N}_l e^{ik_jx^j}e^{ik_0x^0} = \left[ x_0 k_l - x_l \left( \frac{1-e^{-2\ell k_0}}{2\ell} + \frac{\ell}{2} k_m k^m \right) \right]e^{ik_jx^j}e^{ik_0x^0} $$

The main properties of these generators are described, {\it e.g.}, in
Refs.~\cite{majrue,lukieANNALS,gacAlessandraFrancesco,kappanoether,freidkowaNOETHER,nopure}.
Here particularly relevant is the Casimir $\mathcal{C}$,
$$\mathcal{C} = \left(\frac{2}{\ell}\right)^2 \sinh^2 \left( \frac{\ell}{2} P_0 \right) - e^{\ell P_0}P_jP^j ~.$$

Analyses based directly on this construction, such as
the study~\cite{gacmaj}  based on the Casimir and the properties of
the $\kappa$-Minkowski quantum differential calculus~\cite{majidoecklCALC,sitarzCALC},
lead to the conclusion that massless particles (or at least massless waves)
propagate with momentum-dependent speed
$$v = e^{-\ell|\vec{p}|} \simeq 1 - \ell|\vec{p}|~.$$

Before moving on to the next section, where
we review the type of analysis that has been used to challenge this
result, let us set up that discussion by noticing
the following properties of the translation generators
$$[P_l, x_m] e^{ik_jx^j}e^{ik_0x^0} = -i\delta_{lm} e^{ik_jx^j}e^{ik_0x^0} \ ,$$
$$[P_0, x_0] e^{ik_jx^j}e^{ik_0x^0} = ie^{ik_jx^j}e^{ik_0x^0} \ .$$
$$[P_0, x_l] e^{ik_jx^j}e^{ik_0x^0} = 0 \ ,$$
$$[P_l, x_0] e^{ik_jx^j}e^{ik_0x^0} = -i\ell k_l e^{ik_jx^j}e^{ik_0x^0} \ ,$$
where, for the last equation, we used the observation
that $x_0 f(x) = f(x) x_0 + \ell x_j P^j f(x)$,
which follows from (\ref{kappadefCOMMUT})-(\ref{traslCOMMUT}).

\section{$\kappa$-Minkowski phase-space construction}
Our next task is to review the ``$\kappa$-Minkowski phase-space construction"
that has so far provided the only argument against the
momentum dependence of the speed of massless particles
in $\kappa$-Minkowski.

We work at first order in $\ell$
in a 1+1-dimensional $\kappa$-Minkowski spacetime.
This simplified setup helps the clarity of presentation and contains faithfully the hotly-debated issue
concerning the speed of massless particles in $\kappa$-Minkowski.

In the $\kappa$-Minkowski phase-space construction
one describes classical worldlines of particles in terms of an auxiliary
worldline parameter $\tau$
(we denote by $\dot{Q}$ the $\tau$ derivative of an observable $Q$,
so that $\dot{Q} \equiv \partial Q/\partial\tau$).\\
The first ingredient of this derivation of
worldlines is
the following ``$\kappa$-Minkowski Poisson bracket"
for the spacetime coordinates
 \begin{equation}
\left\{ x,t\right\} =-\ell x ~ .
\label{kappadef}
\end{equation}
Then in light of the observations reported at the end of the
previous section one describes the action of translation
generators through the following Poisson brackets
\begin{gather}
\left\{ \Omega ,t\right\} =1,\qquad\left\{ \Omega ,x\right\} =0\ ,\label{timetrasl}
\\ \left\{ P,t\right\} =\ell P,\qquad\left\{ P,x\right\} =-1~.
\label{spacetrasl} \end{gather}
And similarly one relies on the description of boosts here reviewed
in the previous section in order to adopt
the following Poisson brackets between generators of boosts and
of translations:
\begin{equation}
\left\{ \Omega ,P\right\} =0\ ,\qquad\left\{ \mathcal{N}, \Omega \right\} =P\ ,\qquad\left\{ \mathcal{N},P\right\} =\Omega + \ell \Omega^{2} + \frac{\ell}{2}P^{2} ~.\label{poinca}
\end{equation}
It is easy to check that (\ref{kappadef}),(\ref{timetrasl}),(\ref{spacetrasl}),(\ref{poinca})
satisfy all Jacobi identities.
And from (\ref{poinca}) one easily finds the $\kappa$-Minkowski deformed mass Casimir,
\begin{equation}
\mathcal{C}=\Omega^{2}-P^{2}+\ell \Omega P^{2} ~.
\label{casimirOPERATORS}
\end{equation}
One can then use~\cite{jurekvelISOne}
the mass Casimir as Hamiltonian
of evolution of the observables on the worldline of a particle
in terms of the worldline parameter $\tau$.
For this one starts by observing that
Hamilton's equations give the conservation of $P$ and $\Omega$ along the worldlines
\begin{equation}
\dot{P} = \frac{\partial \mathcal{C}}{\partial x} = 0 ~,
 ~~~ \dot{\Omega} = - \frac{\partial \mathcal{C}}{\partial t} = 0 ~.
\end{equation}
We shall denote with $p$ and $E$ these conserved values of $P$ and $\Omega$, and of course
we denote with $m^2$ the conserved value of $\mathcal{C}$, so that in particular
\begin{equation}
m^2=E^{2}-p^{2}+\ell E p^{2} ~.
\label{casimir}
\end{equation}

One then takes into account (\ref{spacetrasl}) in the derivation
of the equations of motion:
\begin{gather}
\dot{t}=\left\{ \mathcal{C},t\right\}
=\frac{\partial\mathcal{C}}{\partial \Omega}\left\{ \Omega ,t\right\}
+\frac{\partial\mathcal{C}}{\partial P}\left\{ P,t\right\} = 2\Omega
-\ell P^{2} \ ,\nonumber \\
\dot{x}=\left\{ \mathcal{C},x\right\}
=\frac{\partial\mathcal{C}}{\partial \Omega}\left\{ \Omega ,x\right\}
+\frac{\partial\mathcal{C}}{\partial P}\left\{ P,x\right\} = 2P-2\ell \Omega P
~. \nonumber \end{gather}

 This evidently implies
 \begin{gather}
t\left(\tau\right)=t_{0}+\left(2E-\ell p^{2}\right)\tau\ ,\nonumber \\
x\left(\tau\right)=x_{0}+\left(2p-2\ell Ep\right)\tau\ ,\nonumber \end{gather}
from which, eliminating the parameter $\tau$ and imposing the Hamiltonian
constraint $\mathcal{C}=m^{2}$, we find
\begin{equation}
x\left(p,x_0,t_0;t\right)=x_{0}+\left(\frac{p}{\sqrt{p^{2}+m^{2}}}-\ell p\left(1-\frac{p^{2}}{p^{2}+m^{2}}\right)\right)\left(t-t_{0}\right)
~. \label{jurekworldlines} \end{equation}
In particular, for massless particles these worldlines give
a momentum-independent particle velocity:
\begin{equation}
x \left(p,x_0,t_0;t\right)=x_{0}+t-t_{0}\ .\label{eq:coordinate velocity}\end{equation}

We shall not dwell much on the setup of this analysis. It suffices to notice
that this is the only argument suggesting that massless particles propagate
with undeformed ($\ell$-independent) speed in $\kappa$-Minkowski. Some of us have elsewhere
argued~\cite{gacMandaniciDANDREA} that analyses such as the one of Ref.~\cite{gacmaj},
which had found evidence of momentum dependence of the speed of massless particles, should
have priority conceptually because they relied on the full quantum structure of the spacetime,
including in particular the noncommutative differential calculus, which instead is moot
in this derivation of worldlines. We can here ignore this debate since we shall find that
when the analysis is advanced to the point of actually establishing 
physically-meaningful
correlations between emission times, detection times and momentum,
rather than merely a ``coordinate velocity", also the derivation of worldlines
based on these ``$\kappa$-Minkowski phase-space constructions"
produces results in  full agreement with the differential-calculus-based analysis of
Ref.~\cite{gacmaj}.

\section{What about Bob?}
The result summarized in the previous section has been known for several years
and was interpreted as a determination of the physical velocity of massless particles
in $\kappa$-Minkowski. In classical spacetime with curvature it is well known that
the coordinate velocity may be affected by coordinate artifacts
 ({\it e.g.} for an observer in
classical de Sitter spacetime  the speed of local photons is
always $1$, but this does not apply to  the coordinate velocity
that observer attributes
to distant photons).
Analogous coordinate artifacts for flat quantum spacetimes were neither expected
nor found until very recently, with the first studies establishing the possibility
of a ``relative locality" in a quantum spacetime.
This was encountered unexpectedly in the two independent
studies\footnote{Before the first results on relativity of locality
 reported in Refs.~\cite{whataboutbob,leeINERTIALlimit}, there had been some papers
arguing about the faith of locality
in the relevant scenarios for spacetime quantization,
but failing to contemplate the possibility that
coincidences of events witnessed by local observers might not appear as such
to observers distant from the events.
In Ref.~\cite{gacIJMPdsrREV} it was observed that two particles on the same
worldline for one observer might not be on the same worldline for another
observer. Similarly in Ref.~\cite{unruh} it was observed that a particle
could be inside a box for one observer but outside the box for another observer.
And a reformulation of such a ``box paradox" for locality was reported
in Ref.~\cite{sabinePRL}, seeking a characterization that would be
suitable for experimental testing.
Ref.~\cite{leePRErelatloc}
argued that the violations of locality discussed in
Ref.~\cite{sabinePRL} could well be in no conflict with known
experimental facts.
Ref.~\cite{gacpiranPRErelatloc} exposed the possibility of sizable violations
of locality in computations involving time dilatation, and
tentatively argued
that there might be logically-consistent relativistic theories
without an absolute objectivity of simultaneity at the same spatial point.}
 reported in
Refs.~\cite{whataboutbob,leeINERTIALlimit},
and then cast into a more satisfactory and more ambitious
conceptual framework in Ref.~\cite{principle}.

We shall here not need much of the broader picture\footnote{We here intend
to contribute to the analysis of
 the standard framework of $\kappa$-Minkowski phase spaces.
In particular,  consistently with the assumptions that are standard
for work on $\kappa$-Minkowski phase spaces, we idealize the processes
of emission and detection of particles.
And we should also stress that, while
one may legitimately argue~\cite{kowaDESITTER} that
some aspects of the structure
of these $\kappa$-Minkowski phase-space constructions
can be described in terms of
the presence of some curvature in momentum space,
the analysis of these $\kappa$-Minkowski phase-space constructions
does not require one to refer directly to the geometry of momentum space.
The formalism recently proposed in Ref.~\cite{principle}
is instead well suited for a description of emission and detection
processes as controlled microscopic particle-physics processes,
and hosts relative locality within a setup which places at center stage
the geometry of momentum space.}
 of the implications and possible formalizations
of relative locality which was recently given in Ref.~\cite{principle}.
For our purposes here it is sufficient to appreciate the
implications of ``relative locality" for coincidences of events,
as described in a picture based on the introduction of spacetime coordinates,
following the rather simple line of analysis advocated in Ref.~\cite{whataboutbob}.
Let us introduce this relative-locality
characterization of coincident events
by drawing a logical line through the history of relativistic theories.
Starting with Galileian Relativity rest is relative (absolute rest requires
a preferred frame, so it is not found in any relativistic theory),
which in turn implies that the quality of being in the same (spatial) position
at different times is not objective.
But in Galileian Relativity simultaneity is still absolute:
when one inertial observer establishes that
two events have the same time coordinate
then all other inertial observers agree.
In Special Relativity distant simultaneity is observer dependent
(``relative simultaneity" or ``relative time"). The relativity of distant
simultaneity in Special Relativity amounts to the fact that
pairs of events which according to one inertial observer are simultaneous
  but occupy different spatial positions (events with some spatial distance between
  them), would not be simultaneous according to other inertial observers.
But in Special Relativity it remains true that
when two events coincide according to one observer they also coincide
according to all other observers: when simultaneity concerns two
events at the same spatial position it remains objective.
This ``absolute locality" of Special Relativity
is in the residual objectivity of simultaneity, which
is subject to
the (spatial) distance between the events but is independent of the distance
between the two events and the observer:
locality is objective
independently of whether it is a ``distant locality" (coincidences of events far from
the observer) or a ``local locality"
(coincidences of events that occur in the origin of the coordinate system
of a given observer).
It is now emerging that in at least certain
quantum spacetimes~\cite{whataboutbob,leeINERTIALlimit,principle}
a weaker form of locality holds:  events that are coincident and occur in the origin
of a given observer Alice are objectively coincident for all observers that
share Alice's origin (all observers which are purely boosted with respect to Alice),
but may not appear to be coincident to observers which are far from Alice.
Locality is still objective, but the abstraction
of ``distant locality" is lost.

The possibility that $\kappa$-Minkowski might be one of these quantum spacetimes
where the objectivity of distant locality is lost was already raised
in Ref.~\cite{leeINERTIALlimit}. We here fully establish
that there is relative locality in $\kappa$-Minkowski,
and we show that, when this is taken into account, 
also within the ``$\kappa$-Minkowski phase-space construction"
one finds that simultaneously-emitted massless particles of different
momentum are not detected simultaneously.

Amusingly a simple way to establish these facts can be based
just on the worldlines rederived in the
previous section. To see this it suffices to
use the setup reviewed in the previous section to formalize
a simultaneous emission occurring in the origin of an observer Alice.
This will be described by Alice in terms of two worldlines, a massless particle
 with momentum $p_1$
and a massless particle
with momentum $p_2$, which actually coincide because of the momentum independence
of the coordinate velocity:
\begin{gather}
x^{A}_{p_1}(t^{A})=t^{A} ~, \label{alicep1} \\
x^{A}_{p_2}(t^{A})=t^{A}~. \label{alicep2}
\end{gather}
It is useful to focus on the case of $p_1$ and $p_2$ such that $|p_1| \ll |p_2|$,
and $|\ell p_1| \simeq 0$
(the particle with momentum $p_1$ is soft enough that it behaves as if $\ell = 0$) while $|\ell p_2| \neq 0$, in the sense
 that for the hard particle the effects of $\ell$-deformation  are not negligible.

A central role in our analysis is played by
the translation transformations codified in (\ref{timetrasl}),(\ref{spacetrasl}).
These allow us to establish how the assignment of coordinates
on points of a worldline differs between two observers connected
by a generic translation ${\cal T}_{a_t , a_x}$, with component $a_{t}$
along the $t$ axis and component $a_{x}$ along the $x$
axis~\cite{jurekvelISOne,arzkowaRelLoc}
\begin{gather}
x'=x-a_{t}\left\{ \Omega ,x\right\} +a_{x}\left\{ P,x\right\} \ ,\nonumber \\
\\t'=t-a_{t}\left\{ \Omega ,t\right\} +a_{x}\left\{ P,t\right\} \ .\nonumber
\end{gather}

Using these we can look at the two Alice worldlines, (\ref{alicep1})
and (\ref{alicep2}), from the perspective
of a second observer, Bob, at rest with respect to Alice at distance $a$ from Alice
(Bob = ${\cal T}_{a , a} \triangleright$ Alice),
local to a detector that the two particles eventually reach.
Of course, since we have seen that the coordinate velocity is momentum independent,
according to Alice's coordinates the two particles reach Bob simultaneously.
But can this distant coincidence of events be trusted?
The two events which according to the coordinates of distant observer Alice are coincident
 are the crossing of Bob's wordline with the worldline of the particle
with momentum $p_1$ and the
crossing of Bob's wordline with the worldline of the particle
with momentum $p_2$.
To clarify the situation we should look at the two worldlines from the perspective
of Bob, the observer who is local to the detection of the particles.


Evidently these Bob worldlines are obtained from Alice worldlines using
the translation transformation codified in (\ref{timetrasl}),(\ref{spacetrasl}).
Acting on a generic Alice worldline $x^A(p^A,x^A_0,t^A_0;t^A)$
this gives a Bob worldline $x^B(p^B,x^B_0,t^B_0;t^B)$ as follows:
\begin{gather}
p^B = p^A - a \left\{ \Omega ,p^A\right\} +a \left\{ P,p^A\right\}
= p^A
 \ ,\nonumber \\
x_0^B = x_0^A - a \left\{ \Omega ,x_0^A\right\} +a \left\{ P,x_0^A\right\}
= x_0^A-a
 \ ,\nonumber \\
\\t_0^B=t_0^A - a \left\{ \Omega ,t_0^A\right\} +a \left\{ P,t_0^A\right\}
=t_0^A-a +\ell a p
  \ .\nonumber \end{gather}
  And specifically for the two worldlines of our interest, given for Alice in
  (\ref{alicep1}) and (\ref{alicep2}),
  one then finds
\begin{gather}
x^{B}_{p_1}(t^{B})=t^{B}-\ell a p_1 \simeq t^{B}
 \ ,\nonumber \\
\\x^{B}_{p_2}(t^{B})=t^{B}-\ell a p_2\ .\nonumber \end{gather}
We have found that,
because of the peculiarities of translational symmetries
of the $\kappa$-Minkowski quantum spacetime,
the two wordlines, which were coincident according to Alice, are
distinct wordlines for Bob.
According to Bob, who is at the detector, the two particles reach the detector
at different times: $t^B \simeq 0$ for the soft particle and
$t^{B} = \ell a p_2$ for the hard particle. And these are the two
massless particles
which, according to the observer Alice who is at the emitter, were emitted simultaneously.

\begin{figure}[h!]
\begin{center}
\includegraphics[width=0.48 \textwidth]{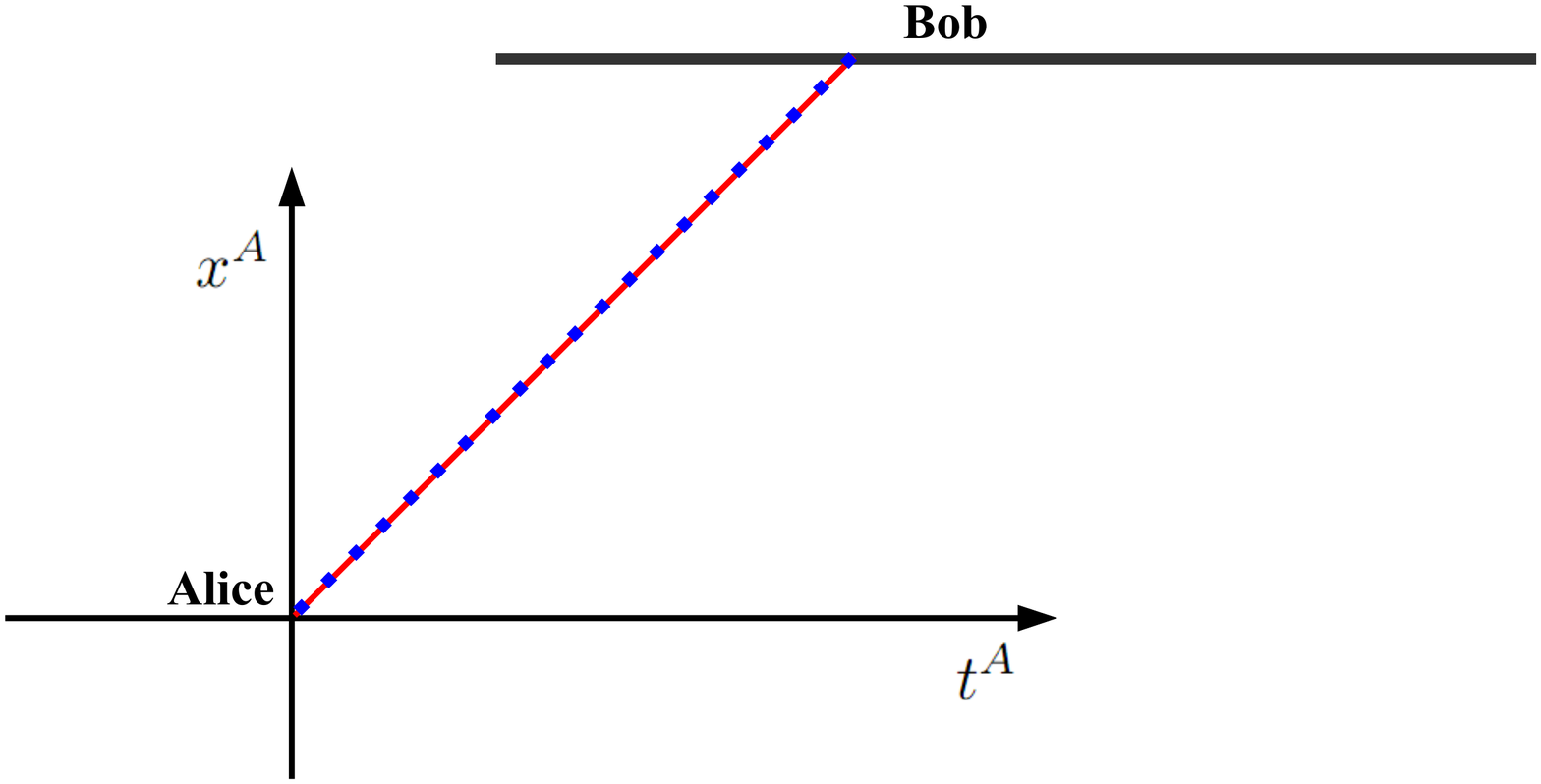}
\includegraphics[width=0.48 \textwidth]{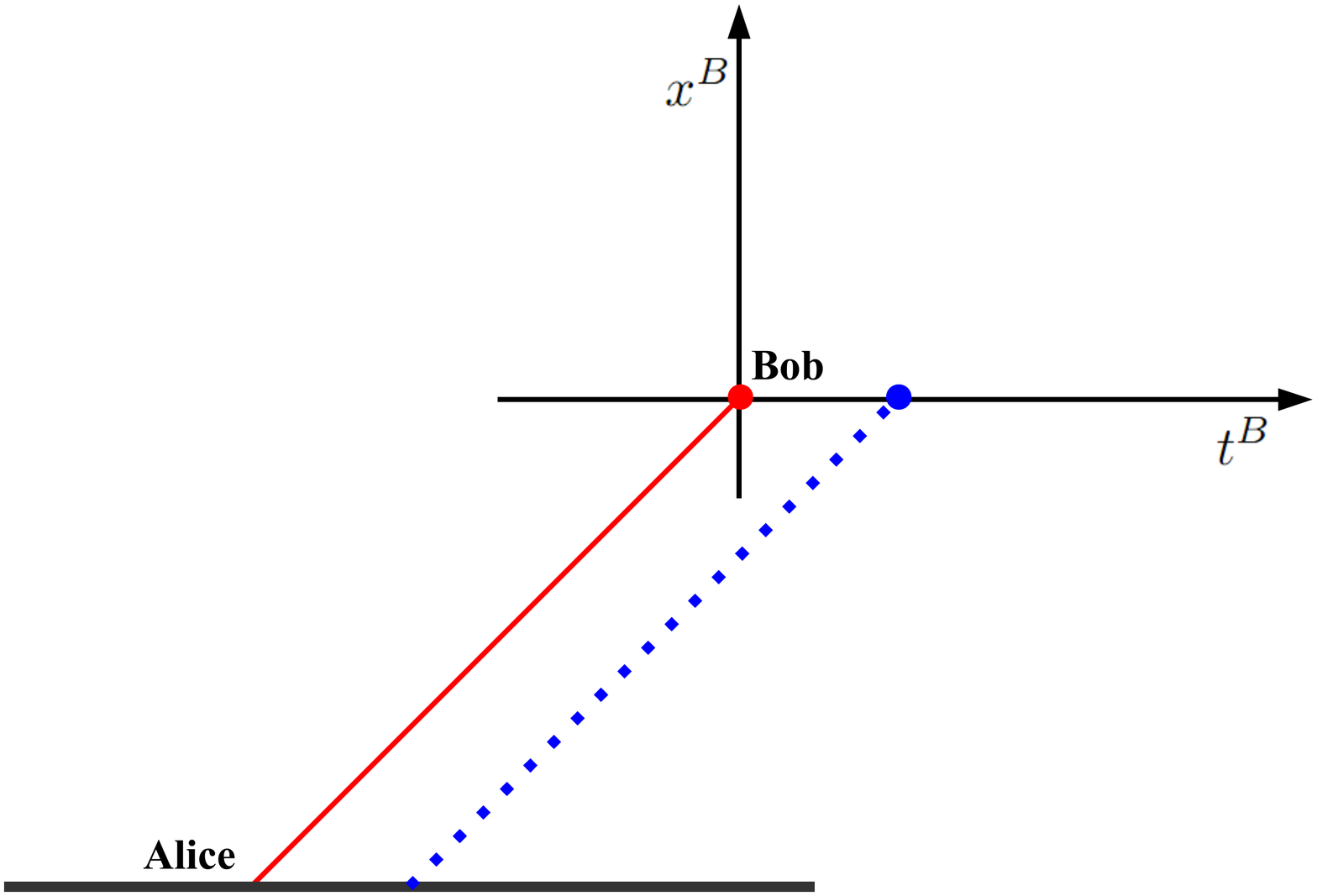}
\caption{Two simultaneously-emitted massless particles
of different momentum in $\kappa$-Minkowski
 are detected at different times.
 The figure shows
 how the simultaneous emission of two such particles and their
 non-simultaneous detection
 is described according to the coordinates of observer Alice (left panel),
 who is at the emitter,
 and according to the coordinates of observer Bob (right panel),
 who is at the detector.}
\end{center}
\end{figure}

Reassuringly this result for momentum-dependence of time of detection
of simultaneously-emitted massless particles,
$$\Delta t = - \ell \Delta p~,$$
is in perfect agreement with the
prediction previously established by a
completely different argument in Ref.~\cite{gacmaj}.

Considering the confused and long history of discussions
of velocity in $\kappa$-Minkowski
it is perhaps worth highlighting the soundness of the operative
procedure by which we have determined this correlation between
momentum of simultaneously-emitted particles
and times of detection.
Our procedure 
rests safely on the robust shoulders of the procedure for
determining the physical velocity measured by inertial observers in classical Minkowski
spacetime, and this connection is allowed by the fact that the properties
of infrared massless particles (properties of massless particles in the infrared limit)
are unaffected by the $\kappa$-Minkowski deformation.
The distant synchronization of the clocks
on emitter/Alice and on detector/Bob is evidently a special-relativistic synchronization,
relying on exchanges of infrared massless particles.
And we implicitly assumed that
the relative rest of Alice and Bob is established
by exchanges of infrared massless particles, so that indeed it can borrow
from the special-relativistic operative definition of
inertial observers in relative rest.
By construction the distance $a$ between Alice and Bob
is also defined operatively just like in special relativity, with the only
peculiarity that Alice and Bob should determine it by exchanging
infrared massless particles.
This setup guarantees that infrared massless particles
are timed and observed
in $\kappa$-Minkowski exactly as in classical Minkowski spacetime.
The new element of the $\kappa$-Minkowski relativistic theory,
concerning the  ``hard" (``high-momentum") massless particles,
then also acquires a sound operative definition by our procedure centered
on the simultaneous emission (with simultaneity prudently established
by the local observer/emitter Alice) of an infrared and a hard massless particle,
then comparing the arrival times at observer/detector Bob (times of arrival
 prudently established according to the local observer, indeed Bob).

In the idealized setting of a sharply flat spacetime our procedure
is applicable for any arbitrarily high value of
the distance $a$. But of course for most realistic applications one will be interested
 in contexts where sharp flatness
of spacetime cannot a priori be assumed, and evidently in such more general cases
the limit $a \rightarrow 0$ of our procedure should be relied upon.
Note however that the characterization of observations (local) and inferences (distant)
given by the coordinates of Alice and Bob, which we summarized in Fig.~1,
evidently remains valid even for small values of $a$: no matter how close Alice and Bob
are, one still has that in Alice's  coordinates the detections at Bob appear
to be simultaneous (while Bob, local to the detections, establishes that they are
not simultaneous) and that
in Bob's  coordinates the emission at Alice appears
to be not simultaneous
(while Alice, local to the emissions, establishes that they are
simultaneous).

\section{Aside on classical Minkowski with noncommutative coordinates}
The robustness of our strategy of analysis
extends even beyond what might be already expected with the observations
we offered so far. Specifically, it is well suited for dealing even with
the most virulent coordinate artifacts.
In this section we provide further evidence of this robustness
by applying our strategy of analysis to a very awkward description of
the familiar classical Minkowski spacetime,
a description that intentionally introduces severe coordinate
artifacts.

We obtain such a description by using
a momentum-dependent redefinition of spacetime coordinates
recently introduced by Smolin in Ref.~\cite{leeINERTIALlimit}:
$$\tilde{x}=x$$
$$\tilde{t}=t + \ell xP e^{- \ell \Omega} \simeq t + \ell xP  $$
 (where the last equality is
valid for $|\ell \Omega| \ll 1$, which is the regime here of interest).\\
This redefinition of coordinates was proposed by Smolin as a way
to probe certain properties of $\kappa$-Minkowski spacetime.
We use the same redefinition in classical Minkowski spacetime,
since it happens to introduce the type
of severe coordinate artifacts that can serve our purposes here.
Postponing some observations on the conceptual implications
of this type of redefinition of coordinates to the next section,
let us now proceed exposing the coordinate artifacts.

So we start by summarizing the properties of the 1+1D
classical Minkowski spacetime,
with its classical Poincar\'e symmetries:
\begin{gather}
\left\{ \tilde{x},\tilde{t}\right\} =0 ~ , \label{classicMink}\\
\left\{ \Omega ,\tilde{t}\right\} =1,\qquad\left\{ \Omega ,\tilde{x}\right\}
 =0 ~, \label{classictimetrasl}
\\ \left\{ P,\tilde{t}\right\} =0,\qquad \left\{ P,\tilde{x} \right\} =-1
 ~,\label{classicspacetrasl} \\
\left\{ \Omega ,P\right\} =0\ ,\qquad\left\{ \mathcal{N}, \Omega \right\} =P\ ,\qquad\left\{ \mathcal{N},P\right\} =\Omega ~,\label{classicpoinca}\\
\mathcal{C} = \Omega^{2} - \Pi^{2} ~.\label{classicCasimir}
\end{gather}
This is the standard setup which famously establishes the momentum independence
of the speed of  massless particles
in classical Minkowski spacetime.

If we now perform
the Smolin redefinition of coordinates, $\tilde{t}=t+\ell xP \, ,~~ \tilde{x}=x$,
the result is a description of classical Minkowski spacetime
in terms of $\kappa$-Minkowski coordinates:
\begin{equation}
\left\{ x,t\right\}
 = -\ell x
~. \label{kappaclassic}
\end{equation}
But then correspondingly the translations of classical Minkowski spacetime
should have a deformed rule of action:
\begin{gather}
\left\{ \Omega,t \right\}
 = \left\{ \Omega,\tilde{t} -\ell\tilde{x}P \right\}
  = 1\ ,\qquad \left\{ \Omega,x\right\} =\left\{ \Omega,\tilde{x}
  \right\} = 0 ~, \label{timetdrunk}
\\ \left\{ P,t \right\} =\left\{ P, \tilde{t} -\ell\tilde{x}P \right\} = \ell P\ ,\qquad  \left\{ P, x \right\} = \left\{ P,\tilde{x} \right\} =-1
 ~. \label{spacetdrunk}
\end{gather}
We can use (\ref{kappaclassic}), (\ref{timetdrunk}), (\ref{spacetdrunk})
and $\mathcal{C} = \Omega^{2} - \Pi^{2}$ to derive the corresponding equations
of motion:
\begin{gather}
\dot{t} = \left\{ \mathcal{C},t\right\} =\frac{\partial\mathcal{C}}{\partial\Omega}\left\{ \Omega,t\right\} +\frac{\partial\mathcal{C}}{\partial P}\left\{ P,t\right\} = 2\Omega - 2\ell P^{2}\ ,\nonumber \\
\\\dot{x}=\left\{ \mathcal{C},x\right\} =\frac{\partial\mathcal{C}}{\partial\Omega}\left\{ \Omega,x\right\} +\frac{\partial\mathcal{C}}{\partial P}\left\{ P,x\right\} =2P\ .\nonumber \end{gather}
{\it i.e.}
\begin{gather}
t\left(\tau\right)=t_{0}+\left(2E - 2\ell p^{2}\right)\tau\ ,\nonumber \\
\\x\left(\tau\right)=x_{0}+2p\tau\ .\nonumber \end{gather}

Therefore the wordlines of particles in classical Minkowski spacetime,
when described in terms of the coordinates $x,t$
take the form:
\begin{equation}
x\left(p,x_{0},t_{0};t\right)=x_{0}+\left(\frac{p}{\sqrt{p^{2}+m^{2}}}+\ell p\frac{p^{2}}{p^{2}+m^{2}}\right)\left(t-t_{0}\right)\ .\end{equation}
which for the
massless case reduces to $x\left(p,x_{0},t_{0};t\right)=x_{0}+\left(1+\ell p\right)\left(t-t_{0}\right)$.
So we have established that the adoption of the noncommutative coordinates $x,t$
produces a coordinate velocity of massless particles in classical Minkowski
spacetime which is momentum dependent.
This must clearly be the result of severe coordinate artifacts introduced
by our awkward adoption of noncommutative coordinates for classical
Minkowski spacetime. So we have here
an ideal opportunity
to test the robustness of our strategy of analysis.
Let us therefore consider a pair of massless particles of different momentum
emitted simultaneously in the origin of observer Alice,
an observer in classical Minkowski spacetime adopting
the noncommutative coordinates $x,t$:
\begin{gather}
x_{p_1}^{A}=\left(1+\ell p_1 \right)t^{A}
~,\label{joc1} \\ x_{p_2}^{A}=\left(1+\ell p_2 \right)t^{A} ~.\label{joc2}
 \end{gather}
Following our strategy of analysis, we are interested in describing the detection
of these two massless particles according to an observer Bob at the detector.
A generic Alice
worldline $x^A(p^A,x^A_0,t^A_0;t^A)$
is mapped into a Bob worldline $x^B(p^B,x^B_0,t^B_0;t^B)$ with
\begin{gather}
p^{B}=p^{A}-a\left\{ \Omega,p^{A}\right\} +a\left\{ P,p^{A}\right\} =p^{A}\ ,\nonumber \\
x_{0}^{B}=x_{0}^{A}-a\left\{ \Omega,x_{0}^{A}\right\} +a\left\{ P,x_{0}^{A}\right\} =x_{0}^{A}-a\ ,\\
t_{0}^{B}=t_{0}^{A}-a\left\{ \Omega,t_{0}^{A}\right\} +a\left\{ P,t_{0}^{A}\right\} =t_{0}^{A}-a+\ell ap\ .\nonumber \end{gather}
In particular Alice's worldlines (\ref{joc1}) and (\ref{joc2})
are mapped into Bob's
\begin{gather}
x_{p_{1}}^{B}\left(t^{B}\right)=\left(1+\ell p_{1}\right)t^{B}
~,\nonumber \\
x_{p_{2}}^{B}\left(t^{B}\right)=\left(1+\ell p_{2}\right)t^{B}\ .\nonumber \end{gather}
So we see that both particles cross the origin of Bob.
As of course expected (in spite of the awkward choice of coordinates)
we have found that in classical
Minkowski spacetime  two
massless particles with different momentum that were simultaneously emitted at Alice
reach Bob simultaneously.

\vskip -0.45cm

\begin{figure}[h!]
\begin{center}
\includegraphics[width=0.44\textwidth]{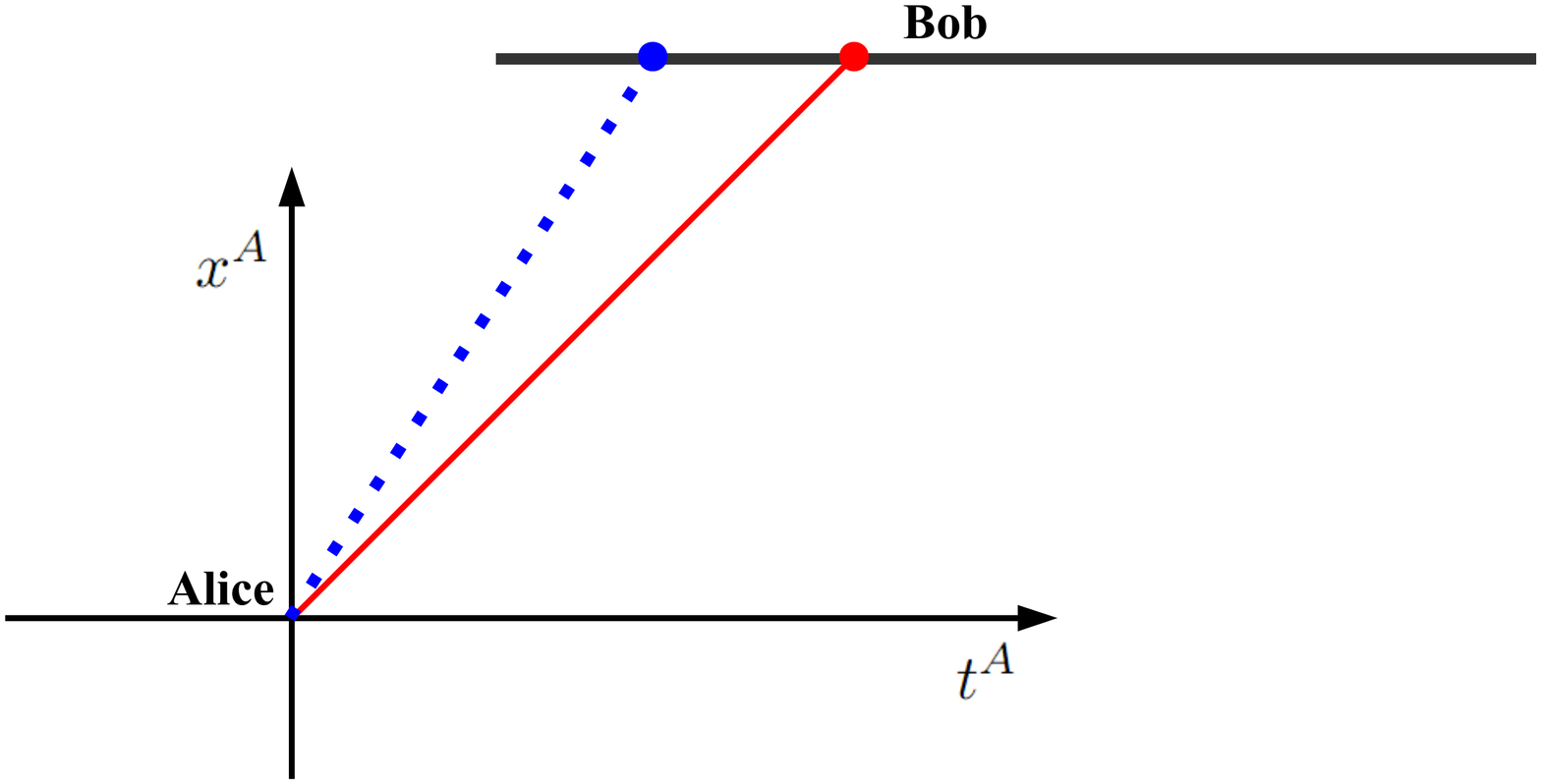}
\includegraphics[width=0.44\textwidth]{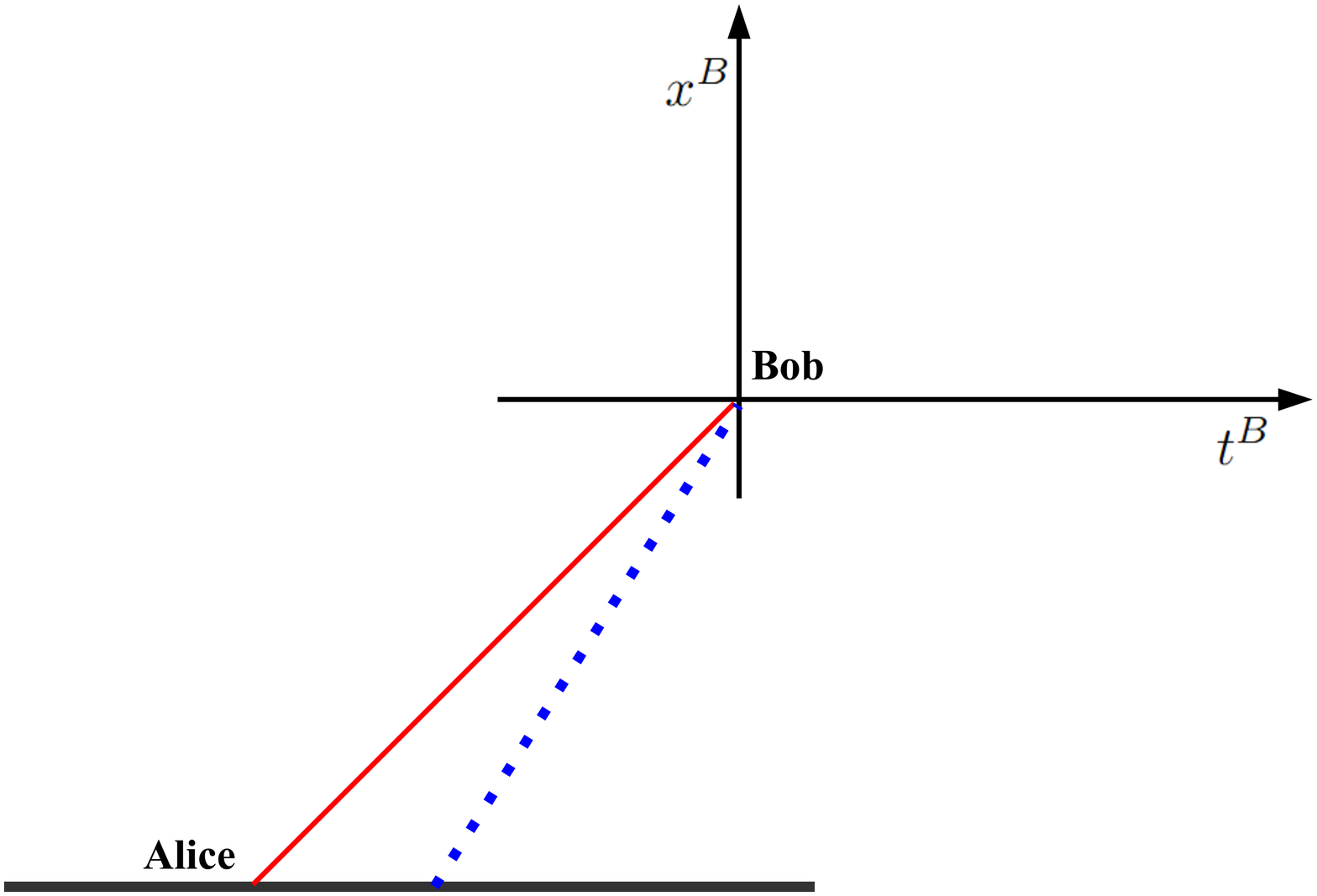}
\caption{Two simultaneously-emitted massless particles
of different momentum in classical Minkowski spacetime
 are detected simultaneously.
 The figure shows, assuming observers
 adopt awkward ``$\kappa$-Minkowski coordinates" for classical
 Minkowski spacetime,
 how the simultaneous emission of two such massless particles and their detection
 is described according to the coordinates of observer Alice (left panel),
 who is at the emitter,
 and according to the coordinates of observer Bob (right panel),
 who is at the detector.}
%
\end{center}
\end{figure}

\section{A first look at other features}

In this section we comment briefly on some other issues that are relevant for the
analysis of $\kappa$-Minkowski theories,
for which our approach appears to be fruitful.
A more in depth investigation of these topics
is the subject of an ongoing investigation~\cite{kappacamilla}.

\subsection{Momentum-dependent redefinitions of coordinates}

From a perspective broader than the confines of $\kappa$-Minkowski
studies perhaps the most striking aspect of our results resides in the
observation that in presence of relativity of locality
the equations of motion (the worldlines, in our case) written by an observer (say, Alice)
are affected by coordinate artifacts. This also has implications for
the idea of what may constitute ``natural" redefinitions of coordinates,
since it removes the main motivation for preferring to restrict one's attention
to redefinitions of coordinates that leave the equations of motion unchanged.\\
Postponing a more detailed analysis of this point to the
forthcoming Ref.~\cite{kappacamilla} let us here offer some related
comments focusing on
the specific example of the change of coordinates for $\kappa$-Minkowski that was
recently proposed by Smolin~\cite{leeINERTIALlimit}, who already for independent
reasons contemplated the possibility of
 momentum-dependent
redefinitions of spacetime coordinates.\\

We actually made use of  Smolin's  redefinition of coordinates
$$\tilde{x}=x$$
$$\tilde{t} \simeq t + \ell xP  $$
in the previous section, for an analysis of classical Minkowski spacetime.
Smolin used~\cite{leeINERTIALlimit}
the same coordinate redefinition as a tool for probing
the structure of theories in $\kappa$-Minkowski spacetime.\\
So one might ask when is this and/or a similar redefinition of
coordinates appropriate and useful?

In relation to this question we should recall that evidently
robust physical features of the relativistic theory
are codified in the readout of
clocks local to the emission of particles
and (appropriately synchronized) clocks local to the detection of particles.
This is something we ended up having to rely upon because the presence
of relative locality spoiled the reliability
(as codifiers of the physical content of the theory)
of the equations of motion.

Momentum-dependent redefinition of spacetime coordinates, such as the
one proposed by Smolin, evidently do not preserve the form of
the equations of motion, but they can in some cases cause no arm to
the readout of clocks local to the emission of particles
and (appropriately synchronized) clocks local to the detection of particles.
We see that Smolin's redefinition of coordinates is an example
of this: the redefinition is moot at $x=0$ so it has no
effect for the times an observer assigns to emission or detection
events that she witnesses as local observer (in her origin with $x=0$).

\subsection{Generalized (Wigner-)Thomas rotations}

Our next task is to characterize
an aspect of the description of the symmetries of $\kappa$-Minkowski
spacetime given in the previous sections in a way that provides a rather
direct link to the relativity of locality and to the possibility
of laws of momentum-dependent transformation of spacetime coordinates.
An attempt to provide an in-depth description of this feature
is ongoing~\cite{kappacamilla}, but we can here sketch out
some preliminary observations.

Within ordinary special relativity
it is well established that there
is an intimate link between
the relativity of simultaneity
and (Wigner-)Thomas rotations~\cite{thomas,wignerthomas,thomasferraro,thomasvisser}.
We want to observe
that there is an analogous link between the type of
relativity of locality found in our $\kappa$-Minkowski setup
and a corresponding generalization of Thomas rotations.

Let us start within ordinary special relativity
in a 3+1-dimensional spacetime (in the 1+1-dimensional case,
whose simplicity we otherwise prefer, there are of course no Thomas rotations).
For our purposes
it suffices to consider two orthogonal
boosts $\mathcal{B}_1$ and $\mathcal{B}_2$
The composition of two such boosts is in general strongly
characterized by the observation that (with $\mathcal{B}_j \simeq \mathbbm{1}+\xi_j\mathcal{N}_j$)
\begin{equation}
\mathcal{B}_2^{-1}\mathcal{B}_1^{-1}\mathcal{B}_2\mathcal{B}_1
=1+\mathcal{B}_2^{-1}\mathcal{B}_1^{-1}\{\mathcal{B}_2,\mathcal{B}_1\}\simeq 1-\xi_1\xi_2\{N_2,N_1\}~.
\end{equation}
In Galileian relativity one has that simultaneity is absolute
and $\{N_1,N_2\}=0$, which produces no Thomas rotation.
In ordinary special relativity $\{N_1,N_2\}= -\epsilon_{123}R_3$,
so that
\begin{equation}
\mathcal{B}_2^{-1}\mathcal{B}_1^{-1}\mathcal{B}_2\mathcal{B}_1
=1-\xi_1\xi_2\{N_2,N_1\}=1-\xi_1\xi_2\epsilon_{123}R_3.
\end{equation}
It is well known that this Thomas rotation $R_3$
is a manifestation of length contraction:
chaining boosts one ultimately produces rotations essentially
because length contraction in one direction (and not in others)
results into an overall rotation of axis.

Of course in ordinary special relativity no rotation of
any sort is produced by chaining translations and boosts
as follows (with $\mathcal{T}=1-a_t\Omega+a_xP$)
\begin{equation}
\mathcal{B}^{-1}\mathcal{T}^{-1}\mathcal{B}\mathcal{T}
=1+\mathcal{B}^{-1}\mathcal{T}^{-1}\{\mathcal{B},\mathcal{T}\}\simeq 1
+ a_t \xi\{N,\Omega\} - a_x \xi\{N,P\} ~,
\end{equation}
focusing for simplicity on the case of a chain in which the spatial part
of translations and the boosts are collinear.
Indeed in ordinary special relativity $\{N_j,P_\alpha\}$ is a
translation, $\{N_j,P_\alpha\} =  \eta_{j\alpha} P_0 - \eta_{0\alpha} P_j$,
and therefore
(focusing again on the case of a chain in which the spatial part
of translations and the boosts are collinear)
one finds
\begin{equation}
\mathcal{B}^{-1}\mathcal{T}^{-1}\mathcal{B}\mathcal{T}
=1
+ a_t \xi P - a_x \xi \Omega ~.
\end{equation}

Postponing as mentioned a more detailed discussion of these issues
to the forthcoming Ref.~\cite{kappacamilla},
let us comment briefly here on how these specific aspects
of ordinary special relativity get modified in our $\kappa$-Minkowski-based
framework with relativity of locality.
First let us notice that
in our (1+1-dimensional) construction we worked at all stages
consistently with the possibility of generalizing all results
to the case of 3+1 dimensions with classical Lorentz sector and
classical space rotations, so we expect that also in $\kappa$-Minkowski
$\mathcal{B}_2^{-1}\mathcal{B}_1^{-1}\mathcal{B}_2\mathcal{B}_1
=1-\xi_1\xi_2\{N_2,N_1\}=1-\xi_1\xi_2\epsilon_{123}R_3$.
In turn this leads us to expect that our $\kappa$-Minkowski-based
framework hosts a rather ordinary mechanism of standard Thomas rotations,
resulting from combinations of boosts.

The key novelty of our relativistic framework with relativity of locality
is found combining boosts and translations.
In fact, if we chain boosts and translations
in $\kappa$-Minkowski we no longer get out a pure
translation
\begin{eqnarray} \mathcal{B}^{-1}\mathcal{T}^{-1}\mathcal{B}\mathcal{T}=\mathcal{B}^{-1}\mathcal{T}^{-1}\mathcal{T}\mathcal{B}+\mathcal{B}^{-1}\mathcal{T}^{-1}\{\mathcal{B},\mathcal{T}\}=
1+\mathcal{B}^{-1}\mathcal{T}^{-1}\{\mathcal{B},\mathcal{T}\} \simeq ~~~~~~~~~~~~
\nonumber\\
\simeq  1+\xi a_t\{N,\Omega\}-\xi a_x\{N,P\}= 1+\xi a_t P-\xi a_x\left(\Omega+\ell\left(\Omega^2+\frac{1}{2}P^2\right)\right) \nonumber
\end{eqnarray}
And evidently the presence of the extra piece $\xi a_x \ell (\Omega^2+P^2/2)$,
which is an element of the universal enveloping algebra,
ultimately produces a ``Generalized (Wigner-)Thomas Rotation", here
defined as a rotation in phase space\footnote{Within the confines of
the $\kappa$-Minkowski framework one finds this explicit link from the
properties of boosts and translations to ``generalized Thomas rotations".
It appears that in general, even setting aside the peculiarities
of the $\kappa$-Minkowski framework, one should expect relative locality
to produce modifications and generalizations of Thomas rotations
and Thomas precessions. In particular, in Ref.~\cite{principle} it was observed
that one can expect, even before specifying anything about boosts,
 that a relative-locality-inducing nontrivial geometry of momentum space
 should provide opportunities for a novel mechanism of "Thomas precession
 in momentum space", a physical effect manifest in contexts where
the evolution of
a system is enclosing a loop in momentum space.}, mixing spacetime and momentum-space coordinates, resulting from chaining
boosts and translations.\\
This is easily verified explicitly:
\begin{eqnarray}
 \mathcal{B}^{-1}\mathcal{T}^{-1}\mathcal{B}\mathcal{T}\rhd x&\simeq& x + \xi a_t \{P,x\}-\xi a_x\left(\{\Omega,x\}+\ell\left(\{\Omega^2,x\}+\frac{1}{2}\{P^2,x\}\right)\right)
 \simeq x - \xi a_t+\ell\xi a_x p\\
\mathcal{B}^{-1}\mathcal{T}^{-1}\mathcal{B}\mathcal{T}\rhd t&\simeq& t + \xi a_t \{P,t\} - \xi a_x\left(\{\Omega,t\}+\ell\left(\{\Omega^2,t\}+\frac{1}{2}\{P^2,t\}\right)\right)
\simeq t - \xi a_x + \ell \xi a_t p - 2\ell \xi a_x E~.
\end{eqnarray}

\section{Closing remarks}

We have here settled a long-standing issue that strongly characterized
the $\kappa$-Minkowski literature.
In light of our results there is now full agreement between the two main
techniques that had been competing for the description of the propagation
of massless particles in $\kappa$-Minkowski. In fact, we found that,
when the relativity of locality is appropriately taken into account,
the analysis of worldlines within the $\kappa$-Minkowski-phase-space setup
reproduces exactly the predictions previously
obtained~\cite{gacmaj} with the alternative technique
based mainly on the properties of
the quantum differential calculus in $\kappa$-Minkowski.

As stressed in our opening remarks,
this agreement among results obtained with different techniques
of analysis of $\kappa$-Minkowski
appears to be significant also 
 for  ``quantum-gravity phenomenology"~\cite{gacLRR}.

We also established here higher standards for the
general objective of construction
of relativistic theories with two non-trivial observer-independent scales,
in the sense of the ``Doubly Special Relativity"
proposal~\cite{gacdsr1,kowadsr,leedsrPRL,dsrnature,leedsrPRD,dsrnature,jurekDSRnew}.
Our set up based on the $\kappa$-Minkowski formalism evidently does
host two such observer-independent scales: the speed-of-light scale
(speed of  photons in the infrared limit, not manifest in our formulas only
because of conventions with $c=1$) and the $\kappa$-Minkowski length/inverse-momentum
scale $\ell$. Particularly for the novel second observer-independent length scale
the ability of mastering relativistic properties is still
rather limited~\cite{gacdsrrev2010}, but we here provided illustrative examples
of several properties not previously contemplated.

We should also stress the role played by relative locality in the observations
 that led us to these results.
 Because of the nature of the context provided by $\kappa$-Minkowski
phase-space constructions
 it turned out to be sufficient to rely on a rather rudimentary perspective
 on relative locality, which took shape
 in Refs.~\cite{whataboutbob,leeINERTIALlimit},
 still centered on properties of spacetime coordinates.
It would be interesting to verify whether the spacetime picture of
these $\kappa$-Minkowski phase-space constructions
could be found to emerge from some given choice of momentum-space geometry
in the sense of the relative-locality framework proposed in
 Ref.~\cite{principle}.

\section*{ACKNOWLEDGEMENTS}

We are very grateful to Michele Arzano,
Francesco D'Andrea,
Laurent Freidel,
Giulia Gubitosi,
Jerzy Kowalski-Glikman,
A.~Marcian\'o,
Flavio Mercati,
Lee Smolin
and Anna Spinelli
for comments on a first draft of this manuscript.
The work of G.~Amelino-Camelia
was supported in part by grant RFP2-08-02 from The Foundational
Questions Institute (fqxi.org).

\end{document}